# Probing quantum confinement within single core-multishell nanowires

*Gema Martínez-Criado[†][*], Alejandro Homs[†], Benito Alén[‡], Juan A. Sans[§], Jaime Segura-Ruiz[†], Alejandro Molina-Sánchez[⊥], Jean Susini[†], Jinkyoung Yoo[◊], Gyu-Chul Yi[◊]*

[†]European Synchrotron Radiation Facility, 38043-Grenoble, France

[‡]IMM, Instituto de Microelectrónica de Madrid (CNM, CSIC), 28760-Tres Cantos, Spain

[§]Department of Applied Physics, Valencia University, 46100-Burjasot, Spain

[⊥]Institute for Electronics, Microelectronics, and Nanotechnology, CNRS-UMR 8520, Department ISEN, F-59652 Villeneuve d'Ascq, France

[◊]National CRI Centre for Semiconductor Nanorods, Department of Physics and Astronomy, Seoul National University, Seoul 151747, Republic of Korea

ABSTRACT: Theoretically core-multishell nanowires under a cross-section of hexagonal geometry should exhibit peculiar confinement effects. Using a hard X-ray nanobeam, here we show experimental evidence for carrier localization phenomena at the hexagon corners by combining synchrotron excited optical luminescence with simultaneous X-ray fluorescence spectroscopy. Applied to single coaxial n-GaN/InGaN multiquantum-well/p-GaN nanowires, our experiment narrows the gap between optical microscopy and high-resolution X-ray imaging, and calls for further studies on the underlying mechanisms of optoelectronic nanodevices.

The controlled growth of core-multishell nanowires allows fundamental investigations of quantum confinement phenomena.[1] So far, sophisticated coaxial band structure engineering has already been used to produce size-dependent effects for advanced light-emitting diodes.[2] Although theory suggests that the carrier distributions exhibit two dimensional confinement in nanowires under a cross-section of hexagonal geometry,[3] its direct observation has never been addressed. Owing to the central role of quantum confinement in limiting carrier dynamics, X-ray excited optical luminescence[4] is accordingly very attractive for imaging single core-multishell nanowires. With the advent of X-ray focusing optics,[5] it has become a challenge to extend the technique into the nanoscale and hard X-ray regime. The emergence of imaging schemes capable of overcoming Abbe's diffraction barrier ($\lambda/2NA$, where $\lambda$ is the wavelength of light and NA is the numerical aperture of the lens) is crucial for optical microscopy.[6] For example, using parallel-detection mode cathodoluminescence-scanning transmission electron microscopy, Lim *et al*. have previously reported direct correlation of structural and optical properties within a single nanowire.[7] Thus, an approach based on more shorter wavelengths than visible light like X-rays represents a key step. In this context, the use of nanometre-sized hard X-ray beam provides unique advantages over near-field imaging approaches (restricted to surfaces with small areas in the region within ~10 nm of the tip or nano-antenna),[8] or far-field imaging techniques (that involve even photoactivatable fluorophores).[9] These advantages are: i) deeper information depths (i.e., without complex sample preparation procedures); ii) temporal resolution; iii) orientational effects by polarization selection rules; iv) site- and orbital-selective, with simultaneous access to K absorption edges and X-ray fluorescence (XRF) emission lines of medium, light and heavy elements, and v) chemical trace sensitivity owing to the high brilliance of the X-ray beam. Overall, hard X-ray excitation can provide simultaneous information about the quantum confinement and chemical composition of the sample as demonstrated here for single coaxial p-GaN/InGaN multi-quantum-well/n-GaN/ZnO nanowires.

Therefore, in this work we show the use of a hard X-ray nanoprobe with optical and chemical contrast mechanisms at the X-ray undulator beamline ID22 at the European Synchrotron Radiation Facility (ESRF)[10] to probe confinement effects within single core-multishell nanowires. Using a pair of Kirkpatrick-Baez Si mirrors, our approach (Fig. 1a) involves collecting X-ray fluorescence (XRF) and XEOL emissions induced by a highly focused and intense hard X-ray beam [60 x 60 nm$^2$ spot size (V x H) with $10^{12}$ ph/s at 29.6 keV] in air at room temperature. The emission of characteristic secondary X-rays is recorded with an energy dispersive Si drift detector, while the luminescence is detected by a far-field optical system. The collected photons are focused on an optical fibre, which transmits the light to a spectrometer equipped with a linear charge coupled device detector. Besides the depth resolution, which is determined by the incident X-ray photons, the spatial resolution of our approach is governed by the spot size of the incident X-ray beam and the diffusion length of carriers.[11] For a quantum emitter, like a single nanowire or quantum dot, the resolution can be enhanced several times compared with standard optical excitation. In combination with the high brightness of the third-generation synchrotron source, our current diffraction limited X-ray lenses[12] allow for nanoscale lateral and spectral analysis with short acquisition times (typically from 200 ms to 1 s per spectrum).

As a proof of concept, we apply this technique to single coaxial p-GaN/InGaN multi-quantum-well/n-GaN/ZnO nanowires grown by metal-organic vapour phase epitaxy.[13] Recent calculations indicate that the modulation of the radial elemental composition within nanoscale hexagonal geometry introduces new complexities that create novel confinement effects.[3,14] The heterostructure consists of an inner ZnO nanotube core (~500 nm diameter), and sequentially deposited n-type GaN (200 nm), three periods of GaN/In$_{0.24}$Ga$_{0.76}$N multi-quantum-wells (MQWs) (2-nm-thick well and 13-nm-thick barrier), and a p-type GaN shell (120 nm).[15] A schematic cross-section of the sample and its energy band alignment is shown in Fig. 1b and c, respectively. The n-type GaN inner and the p-type GaN outer shells serve as electron and hole injection layers, respectively. The InGaN provides a band gap energy well for efficient trapping and radiative recombination of injected carriers, and the p-n interface extends along the entire length of the nanowire with carrier separation in the radial direction. In this geometry, photogenerated carriers can reach the junction with high efficiency since diffusion lengths are relatively short. This architecture exhibits two notable advantages over planar group-III nitride structures in terms of optical performance. The first advantage is the reduced lattice mismatch between GaN and ZnO (~1.86 %) results in low interface strain, and low dislocation and stacking fault densities, thus reducing nonradiative recombinations. The second advantage is that it should not suffer from large internal polarization-related electric fields, which could strongly separate the electron and hole wavefunctions, reducing the probability of radiative recombination.[16]

Earlier optical studies of an ensemble of such core-multishell nanowires have highlighted the strong green and blue emissions at room temperature coming from efficient radiative recombinations in the quantum well structures.[13] In this work, thanks to the spatial resolution of the X-ray spot, these recombination channels are spatially resolved in the radial direction of single nanowires and correlated with compositional analyses. We collected side-to-side XRF and XEOL data at $\theta = (15 \pm 5)°$ with respect to the sample surface. Figure 2 shows the nano-XRF results obtained by scanning the nanowire along the radial direction. The measurements identify the precise spatial location of the nanowire and reveal several points: The three-dimensional (3D) representation of the XRF data shows that the major elements exhibit high contrast, indicative of uniform core-multishell encapsulation in the radial chemical composition consistent with the targeted heterostructure (Fig. 2c-e); Within the sensitivity of our experimental setup, the sharp elemental profiles suggest that no interdiffusion worthy of comment took place across individual shells (Fig. 2b); The average XRF spectrum reveals unintentional dopants (e.g., Cr, Fe and Ni) which are probably incorporated during the initial pattern formation and position-controlled selective MOVPE growth.[13,15] The corresponding XRF maps indicate a homogeneous incorporation of these impurities at the length scale of the beam size (data not shown here). In summary, the XRF findings reflect an uniform radial growth process, without relevant signatures of deposition-induced diffusion and/or agglomeration effects.

Despite the homogeneous elemental distributions, the hexagonal geometry can lead to additional quantum size effects and therefore to inhomogeneous carrier distributions. This is revealed by XEOL data simultaneously recorded in the same nanowire. First, the XEOL spectra obtained from different excitation positions (Fig. 3a on linear scale) showed different spectral shapes, indicating the inhomogeneous distribution of the radiative recombination channels responsible for the visible transitions. The spectra exhibit three principal emission bands: a shoulder at 2.25 eV attributed to the common yellow band from point defects in GaN (probably Ga vacancies), a dominant green line at 2.45 eV attributed to the transitions from the $In_{0.24}Ga_{0.76}N$/GaN MQWs, and a weak blue peak at 2.85 eV attributed to the band-to-acceptor emission of the outmost Mg-doped GaN layer.[13] Second, the 3D projection of the emissions located at 2.25 and 2.85 eV reveals a uniform distribution of gallium vacancies and impurities in Fig. 4c and d, in good agreement with the XRF results. In contrast, the 3D illustration of the band emitted at 2.45 eV indicates that the InGaN-related emissions radiate mainly from the six corners of the hexagon (Fig. 3c-e). The latter characteristic strongly suggests the existence of additional carrier confinement effects.

On a pixel-by-pixel basis, a multiple-Gaussian fitting procedure was applied to each XEOL spectrum using PyMca code.[17] The resulting spatial distribution of the InGaN-related transition energy, linewidth and intensity can thus be revealed (Fig. 4). To obtain further information on these radiative recombinations, we recorded the XEOL spectrum for various photon fluxes (Fig. 4a on logarithmic scale). Although the spectral shapes do not change much even for three orders of magnitude difference in photon flux, the relative intensities grow faster for the InGaN-related emission at 2.45 eV with a reduction of its linewidth (clearly observed in linear scale). In addition, there is a small but measurable blue energy shift that results from the screening of the polarization electric field in the slightly strained MQWs structure. XEOL images and spectra show that the InGaN-related emission maxima, line shapes and intensities are nearly identical at the hexagon corners. The XEOL spectra indicate that emission maxima systematically shift to higher energy with a reduced linewidth. Although previous luminescence studies suggest that apparently broad linewidths can be attributed to small QW thickness fluctuations and/or alloy broadening mechanisms,[18] our results signify less spectral diffusion resulting from Stark shifts produced by changing local electric fields at the hexagon corners.[19] The data allowed us to separate two different mechanisms responsible for the observed effects: (i) the presence of a small quantum-confined Stark effect and (ii) radial confinement effect.

The combination of spatially resolved XEOL and coaxially grown quantum structure additionally reflects the carrier diffusion in the GaN barrier. At room temperature, we measured the XEOL spectra as a function of the X-ray beam position along the nanowire radial axis with a 50 nm spacing between consecutive X-ray beam spots. As shown in Fig. 4a, the excitation position-dependent XEOL spectra change in spectral shape and intensity. Thus, the ambipolar diffusion length, $L_d$, can be roughly estimated by the integrated XEOL intensity of the InGaN-related emission as a function of the X-ray beam position,[20] $I_{MQW} = I_0 \exp(-x/L_d)$, where $I_{MQW}$ is the XEOL intensity of the transition from the InGaN MQW, $x$ is the distance between the X-ray beam position and the MQW region, and $I_0$ represents a scaling factor. From the fit of XEOL intensity in Fig. 5b (black curve), the ambipolar diffusion length in the radial direction of the coaxial p-GaN/InGaN multi-quantum-well/n-GaN/ZnO nanowire is about $(150 \pm 20)$ nm at room temperature. Compared to the values published so far [200-600 nm for InGaN/GaN,[21,22] the diffusion length estimated may imply larger mobility-limiting scattering mechanisms.

Position-dependent changes in XEOL spectra are caused by the Stark shift that arises due both to the small piezoelectric field and to spontaneous polarization. In addition to these effects there are changes attributable to the enhanced carrier confinement within the MQWs merging at the corners; this interpretation is supported by simplified theoretical calculations.[1] The single-particle Schrödinger equation for the conduction band states (electrons) can be written as:

$$\left(-\frac{\hbar^2}{2m_c}\nabla^2_{x,y}+V(x,y)\right)\Psi(x,y)=E\Psi(x,y)$$

where $\nabla^2_{x,y}$ is the Laplacian for the x,y coordinates, $m_c$=0.20 is the GaN electron effective mass, and $V(x,y)$ the potential. The results for the valence band states (holes) are analogous (only the effective mass changes) to those for electrons. As we consider an infinite nanowire, the axial coordinate can be omitted. To simplify the calculation, we assumed that the nanowire heterostructure consists of a hexagonal shell of InGaN surrounded by GaN. Due to the thickness of the GaN barriers (13 nm), our model reproduces fairly well the sample structure. Therefore, the potential $V(x,y)$ has been defined in a square box, assuming infinite barriers along the borders. Inside this bidimensional box, the potential takes the value of $E_b$ = 3.5 in the GaN region and $E_b$ - Δ in the InGaN region, where Δ = 0.40 is the offset between GaN and InGaN. The strain effects on the potential profile have been checked by finite element calculations and only induce a slight change in the band offset, without substantial modification of the potential profile. Thus, the Schrödinger equation has been solved with the finite differences method and the results are summarized in Figure 5. We have represented the square of the electron wave function of the lowest-energy state of the conduction band for three well sizes, together with a three dimensional representation of the InGaN-related emission measured by XEOL from three different nanowires. As expected, there is a confinement of the wave functions preferentially at the hexagon corners, owing to the symmetry imposed by the shape of the nanowire. Moreover, as the well dimension decrease, the confinement effect becomes larger. Compared to the calculations published by B. M. Wong *et al.,*[3] here no polarization charges are assumed, and the potential profile is modelled by a step function, a reasonable approximation for the case of a quantum well.[23] The most important point is that a higher confinement of the wave function at the hexagon corners is observed. In remarkable consistency with the XEOL imaging, the statistical significance of these spatially-dependent data is corroborated by the measurements carried out on several independent nanowires (Fig. 5 c-e). In short, our results show the appearance of a finite charge separation over the hexagon, which ends in an overall greater overlap between the wave functions of electrons and holes at the corners.

In summary, the observation of the carrier confinement effects under hexagonal cross-section in single core-multishell nanowires presented here gives a glimpse of the new research directions such a hyperspectral imaging method can provide. It represents a step towards not only the validation of theories of quantum confinement for nanodevices, but also the realization of nanostructures with spectroscopic properties that could prove advantageous in light-emitting diodes. Its great potential becomes more valuable when time resolving power is added, as well as when this technique is used in conjunction with other methods, such as X-ray absorption spectroscopy and X-ray diffraction. Though readily accessible with current X-ray optics, the technique will benefit from the high numerical aperture of next X-ray lenses produced by state-of-the-art fabrication methods and also from the high-energy synchrotron radiation facilities, where the same type of analysis could be performed with lower emittances by using longer beamlines and even shorter wavelengths, thus increasing both the penetration depth and lateral resolution.

FIGURES

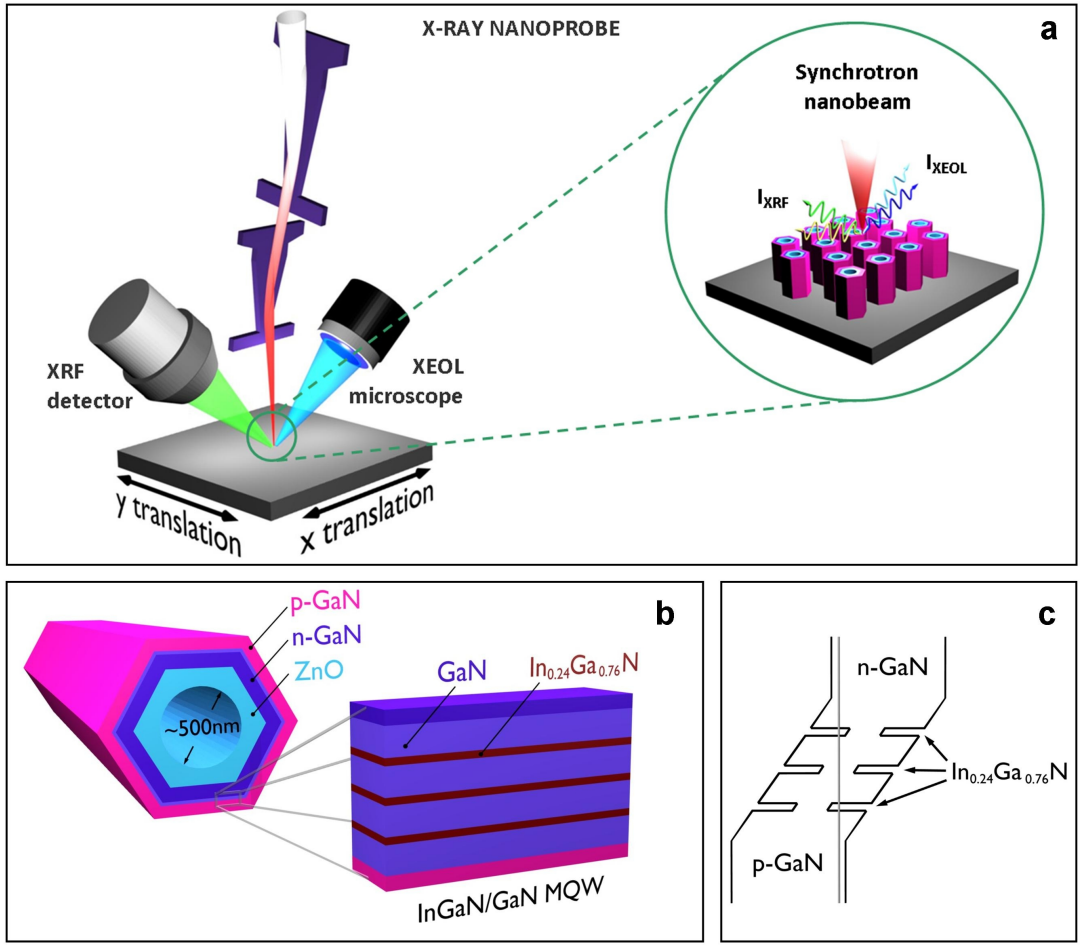

**Figure 1.** (a) Schematic of experimental setup. The X-ray nanobeam impinges on the sample, which emits luminescence and X-ray fluorescence photons. X-ray induced optical luminescence and X-ray fluorescence spectra are recorded with a far-field optics system and a Si drift detector, respectively, for each raster position of the specimen. (b) Schematic of the GaN/In1-xGaxN/GaN/ZnO nanowire heterostructure and magnified cross-sectional view highlighting the GaN/InGaN MQW. (c) Energy band diagram with a dashed line indicating the position of the Fermi level.

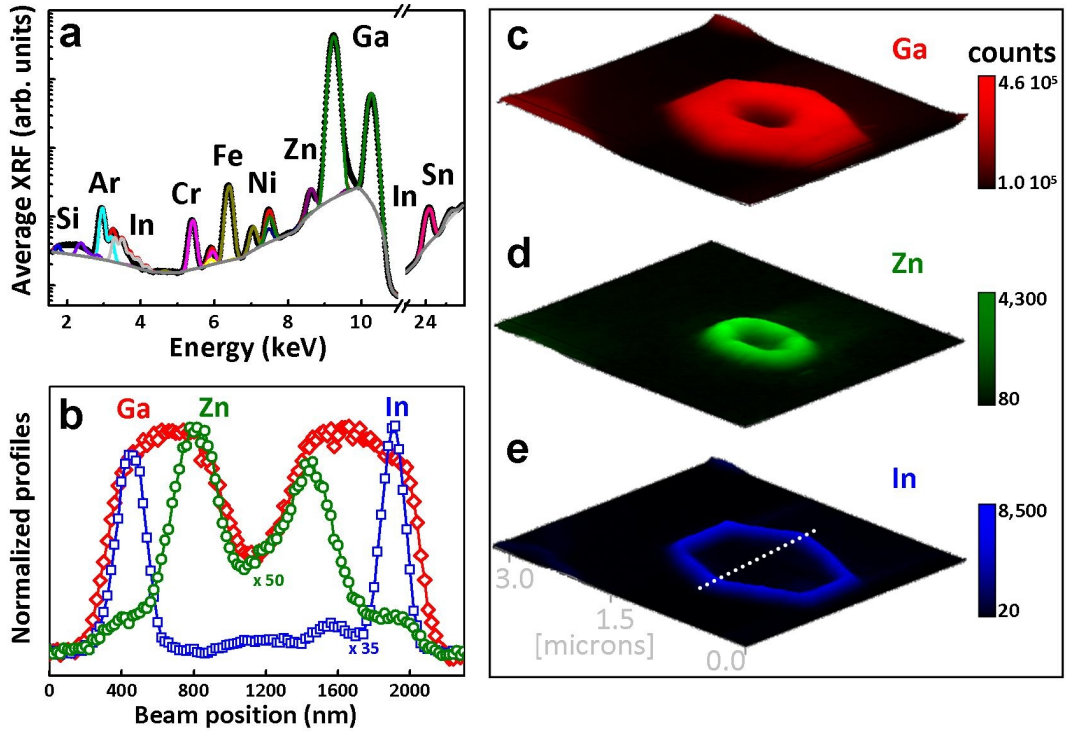

**Figure 2.** (a) Average XRF spectrum recorded over a 2 × 1.4 µm2 cross-sectional scan of an individual nanowire. (b) Normalized XRF line profiles for Ga (red symbols), Zn (green symbols), and In (blue symbols), respectively, collected along the white dotted line with 20 nm step size. (c, d, e) Elemental mapping of the same nanowire, indicating spatial distribution of Ga (red), Zn (green), and In (blue), respectively. Map size: 2 × 1.4 µm2; pixel size: 50 nm; counting time: 1 s/point.

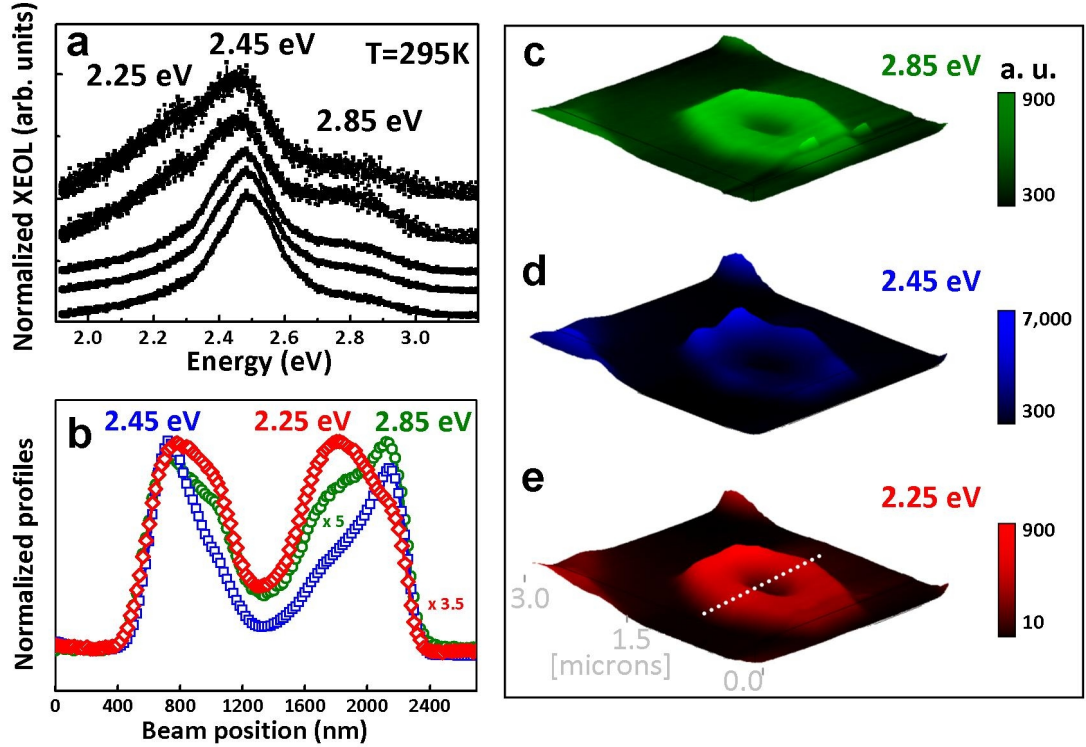

**Figure 3**. (a) XEOL spectra collected at different positions across the nanowire. The dominant green line at 2.45 eV is attributed to the radiative transitions from the $In_{0.24}Ga_{0.76}N$/GaN MQWs, the weak blue peak at 2.85 eV is associated to the band-to-acceptor emission of the outmost Mg-doped GaN layer, and the shoulder at 2.25 eV is assigned to the common yellow band from GaN. (b) Normalized XEOL line profiles for the transition at 2.25 eV (red symbols), 2.45 eV (blue symbols), and 2.85 eV (green symbols), respectively, collected along the white dotted line with 20 nm step size. (c, d, e) Luminescence mapping of the same nanowire, indicating spatial distribution for 2.25 eV (red), 2.45 eV (blue), and 2.85 eV (green) bands, respectively. Map size: $2 \times 1.4$ μm2; pixel size: 50 nm; counting time: 1 s/point.

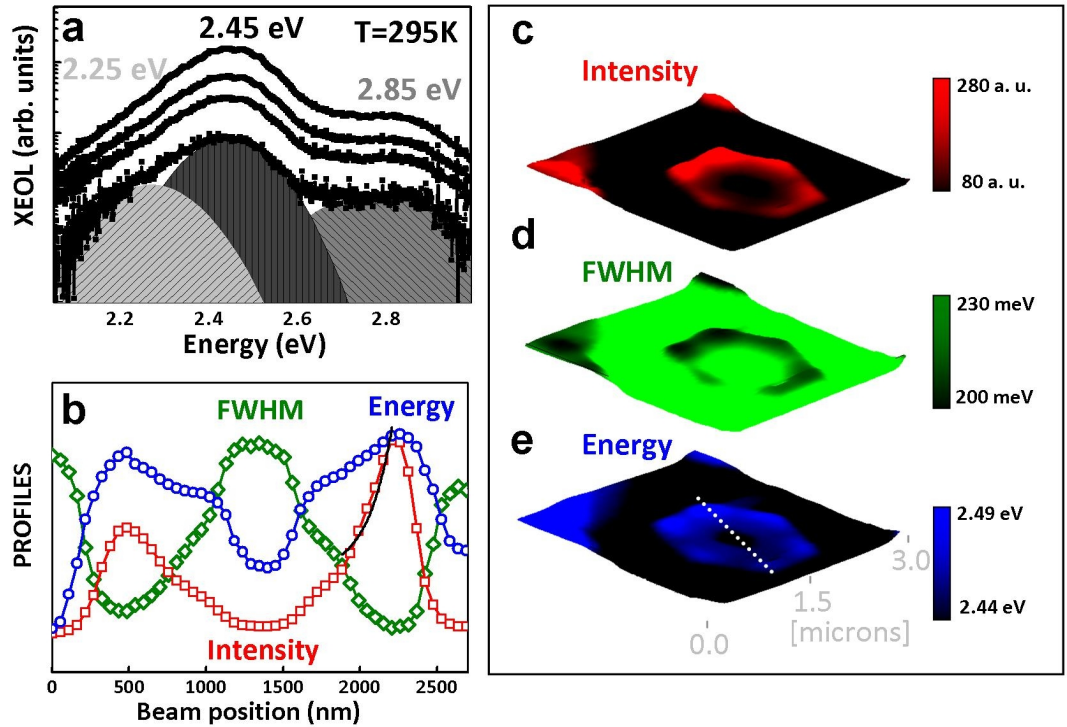

**Figure 4**. (a) Average XEOL spectra recorded from an individual nanowire with photon fluxes ranging from 1010 to 1012 ph/s. (b) Normalized line profiles resulting from the spectral decomposition of the dominant transition at 2.45 eV: Intensity (red), FWHM (green), and energy (blue) along the white dotted line with 50 nm step size. The black curve corresponds to an exponential decay fit performed to deduce the carrier diffusion length. (c, d, e) Spatial distribution for the exciton integrated intensity (red), FWHM (green), and energy (blue).

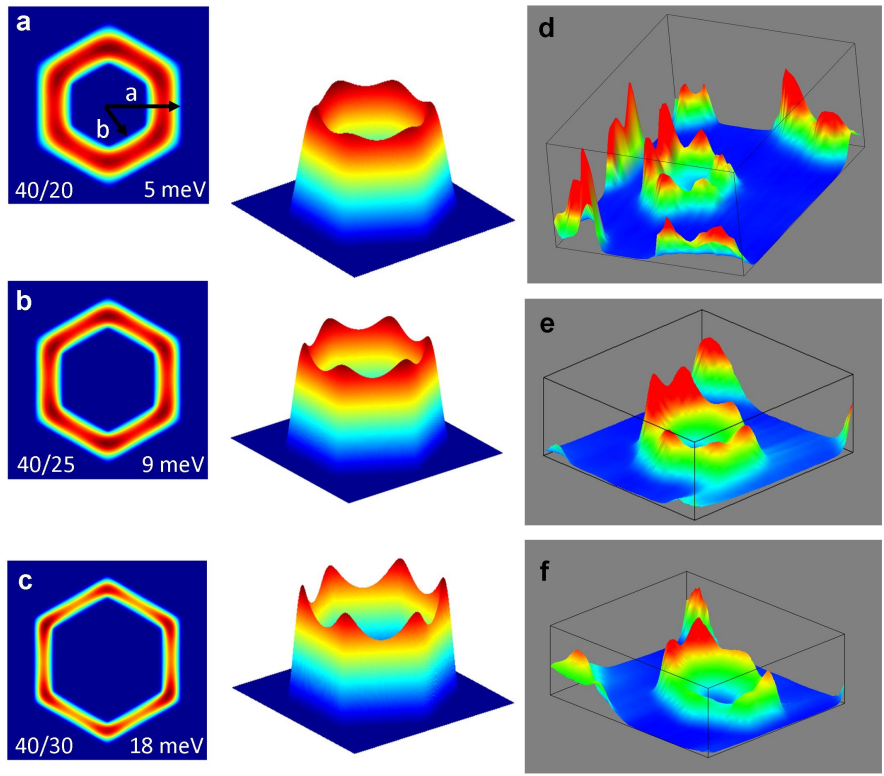

**Figure 5**. (a, b, c) Calculated square of the electron wave function of the lowest-energy state of the conduction band, including a 3D representation. The a/b ratios used in the simulation are included together with the energy (in meV) of the electron wave functions (referred to the conduction band of the quantum well). (d, e, f) Spatial distribution of the InGaN-related recombination at 2.45 eV collected from three different nanowires.

AUTHOR INFORMATION

**Corresponding Author**

*E-mail: gmartine@esrf.fr.

**Notes**

The authors declare no competing financial interest.

ACKNOWLEDGMENT

The authors thank Irina Snigireva and Armando Vicente Solé for their assistance with the SEM measurements and data processing using PyMca, respectively. We thank Rémi Tocoulou and Peter Cloetens for their help and the ESRF for the beam time allocated. We also thank Andrei Rogalev for the valuable discussions and Gary Admans for the critical reading of the manuscript. This work has been partially supported by the NANOWIRING Marie Curie ITN (EU project no. PITN-GA-2010-265073).

**Table of Contents (TOC) Image**

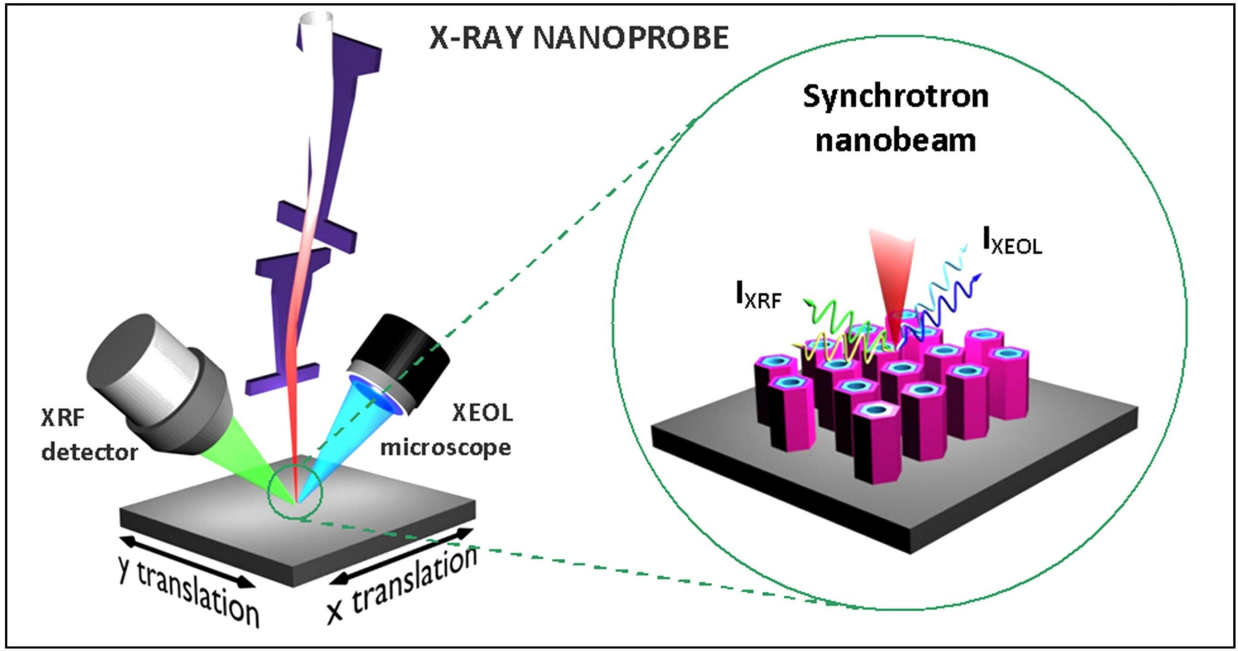
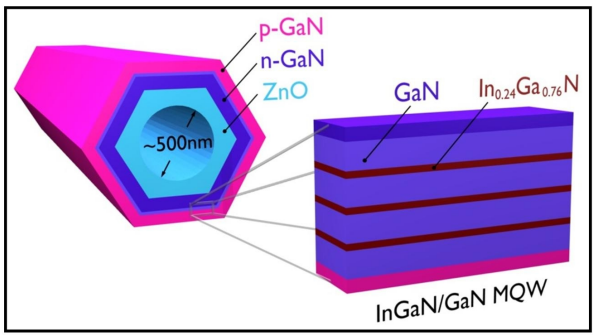
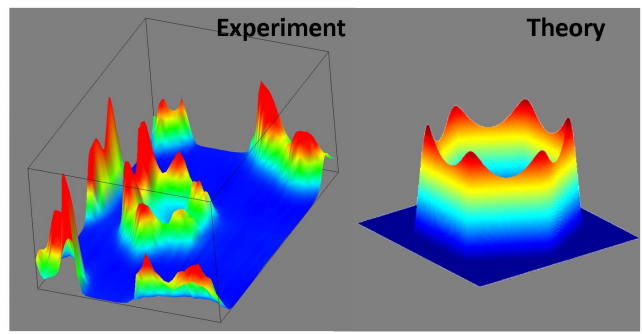